\documentclass[aps,prl,twocolumn,epsfig]{revtex4}

\usepackage{graphicx}

\begin{document}

\title
{Delay of Disorder by Diluted Polymers}
\author{A.V.~Kityk and C.~Wagner $^{2,*}$}
\affiliation{ $^1$ Institute for Computer Science, Technical
University of Czestochowa, Electrical Eng. Dep., Al. Armii
Krajowej 17, PL-42200 Czestochowa, Poland  \\ $^2$
Experimentalphysik, Universit\"at des Saarlandes, Postfach 151150,
66041 Saarbr\"ucken, Germany    $^*$ c.wagner at
mx.uni-saarland.de}

\begin{abstract}
We study the effect of diluted flexible polymers on a disordered
capillary wave state. The waves are generated at an interface of a
dyed water sugar solution and a low viscous silicon oil. This
allows for a quantitative measurement of the spatio-temporal
Fourier spectrum. The primary pattern after the first bifurcation
from the flat interface are squares. With increasing driving
strength we observe a melting of the square pattern. It is
replaced by a weak turbulent cascade. The addition of a small
amount of polymers to the water layer does not affect the critical
acceleration but shifts the disorder transition to higher driving
strenghs and the short wave length - high frequency fluctuations
are suppressed.
\end{abstract}

 \maketitle It is
known from the $1950$s \cite{Toms} that small amounts of flexible
polymers can cause drag reduction in turbulent flow. But both
experimental and theoretical investigations have not yet
conclusively answered the questions what the fundamental
mechanisms of drag reduction are and statistical  and stability
arguments coexist \cite{Gyr}. A difficulty in the theoretical
description of drag reduction in fully developed turbulence (e.g.
in pipe flow) is that the polymer effect, the drag reduction,
occurs in a range where any theoretical treatment of the flow must
be fully nonlinear. So called \textit{weak turbulence}
\cite{zakharov}, reported for different systems like semiconductor
lasers an waves, obeys similar properties like fully developed
turbulence, e.g. the energy is injected on large scales and
transported via a continuous cascade to shorter, viscous length
scales. But typical Reynolds numbers are much lower and for the
analytical treatment small background waves can be taken as the
small parameter around which a linearisation can be performed.

\begin{figure}
\includegraphics[ width=0.95\linewidth]{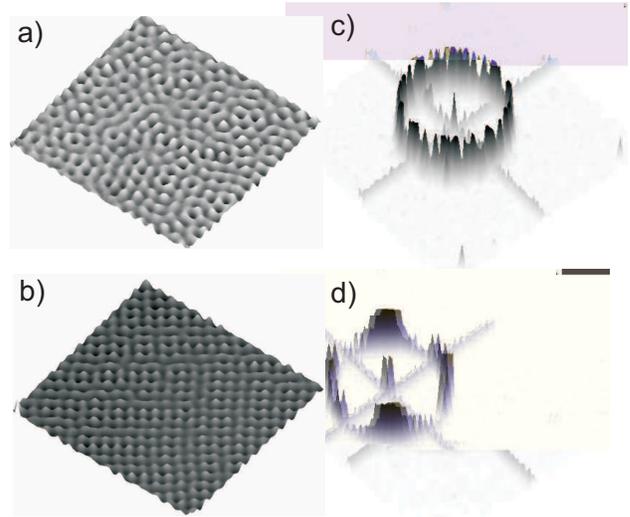}
\vspace{0.03cm} \caption{Snapshots of the interfacial waves at a
driving amplitude $a_0=31.7$ m/s$^2$. The lower fluid layer is (a)
the pure dyed water sugar solution and (b) with 110 ppm PEO. (c)
and (d) are the corresponding power spectra. \label{fig1} }
\end{figure}

In this Letter we present experimental measurements that show how
polymers can affect the transition to a weakly turbulent state and
delay the disordering. We present quantitative data of
parametrically driven waves (Faraday waves, see Fig. \ref{fig1})
at the interface of two liquid layers.
Due to the low viscosities the system evolves already at low
driving strengths from the square state into disorder. Our
measurements show that by adding even of a small amount of
flexible polymers (Polyethylenoxide, PEO) to the aqueous layer the
transition into disordered state occurs at considerably higher
driving accelerations whereas the critical acceleration of the
primary bifurcation remains unaffected. We will see that the
pronounced elongational viscosity of the polymer solution accounts
for this behavior.

The experimental setup is described in detail elsewhere
\cite{kityk}. In order to overcome the problems associated with
light refraction at the interface of a free surface we chose to
work with a system of two layers of refractive index matched
unmiscible liquids of different densities. The replacement of air
as the upper fluid for \textit{free} surface waves is
hydrodynamically nothing but an increase in viscosity and density,
but the actual interfacial deformation can be simply determined by
the absorption of light passing the lower, dyed layer. The method
has already been proven to allow for quantitative determination of
the complete spatio-temporal spectrum of ordered wave states, with
a resolution in amplitude better than one percent of the maximum
wave height. The lower liquid is a sugar-dye-water mixture
(density $\rho=1200 kg/m^3$, viscosity $\eta = 3.2$ mPas) in which
the high molecular PEO polymer ($Mw = 6 \times 10^6 amu$) can be
dissolved. The highest polymer concentration is $110$ ppm well
below the critical overlap concentration $c^\star = 300$ ppm. The
upper liquid is a silicon oil (density $\rho = 910 kg/m^3$,
viscosity $\eta = 0.65$ mPas). The low viscosity was chosen to
minimize the influence of the upper layer, similar like in free
surface waves. The interfacial tension is $\sigma = 35$ dynes/cm,
yielding a capillary length of $\ell_c=(\sigma/(\Delta \rho
g)^{1/2}= 3mm$, with $g$ the earth acceleration and $\Delta \rho$
the differences in the fluid densities. The thickness of the water
sugar (silicon oil) layer is $h_0 = 3mm$ ($7mm$). A closed
circular vessel with glass plates at the top and bottom is
attached to an electromagnetic shaker. The temperature is hold
constant at $T = 23 \pm 0.1 ^\circ$C.  The system is accelerated
in the form $a(t)=a_0 sin(\Omega t)$ and the driving frequency is
always $\Omega/2\pi = 47$Hz.  We are in the regime of the generic
subharmonic response of the Faraday Experiment and the waves
oscillate with a fundamental frequency $\omega_0=\Omega/2$. It is
related via the dispersion relation to a wavelength $2 \pi/k_0 = 5
mm$ and we are close to the infinite depth limit $tanh(k_0h) = 1$.
Our measurements are performed by increasing the driving strength
$a_0$ in steps of $2\%$. After an equilibration time of $60$
seconds the interface is filmed for two seconds by an 8-bit camera
with a spatial resolution of $256 \times 256$ pixels at repetition
rate of $500$ Hz. The maximum acceleration is $a_0 = 95 m/s^2$ and
for $a_0 > 54 m/s^2$ first droplet ejections occur, but only for
$a_0
> 70 m/s^2$ the interface disintegrates significantly.

\begin{figure}
\includegraphics[ width=0.85\linewidth]{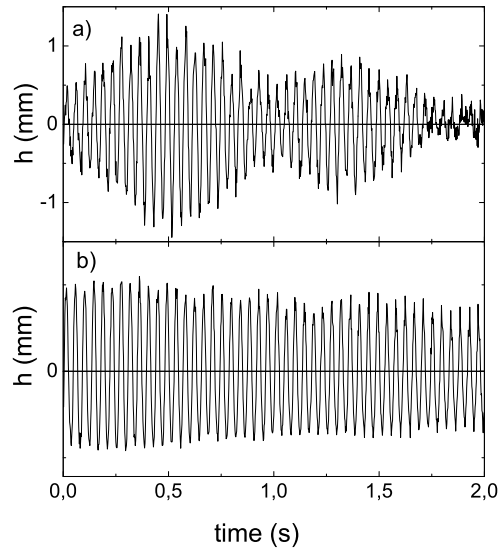}
\vspace{0.03cm} \caption{A typical temporal evolution of the
interfacial deformation at an arbitrary point near the center of
the container. Parameters are like in Fig. \ref{fig1}.\label{fig2}
}
\end{figure}

All solutions show the same critical acceleration $a_c=17.5$
m/s$^2$ at which the interface becomes unstable first.  Near onset
a pattern of squares is formed. Figure \ref{fig1} shows a typical
snapshot of a measurement at a driving amplitude of $a_0=31.7$
m/s$^2$ with the lower layer (a) the water sugar solution and (b)
after the addition of 110 ppm PEO. Strikingly, for the diluted
polymer solution the quadratic order is still clearly to identify
while the order in the water sugar solution is already "melted". A
typical temporal realization of the interface displacement
$h(x,y,t)$ is shown in Fig. \ref{fig2}. For the polymer solution
we observe a small drift while the water sugar solution is
strongly intermittent.

\begin{figure}
\includegraphics[ width=1.1\linewidth]{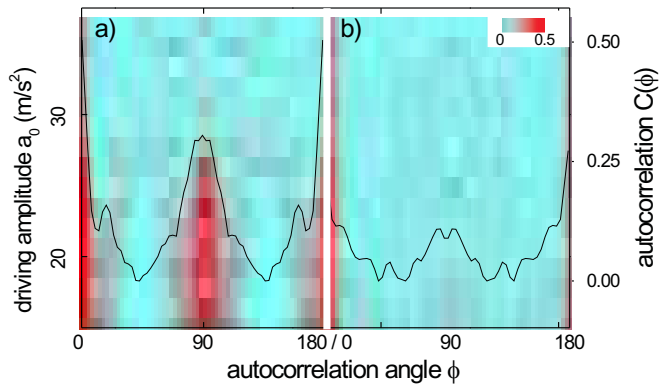}
\vspace{0.03cm} \caption{(Color online:) The temporal averaged and
baseline subtracted angular autocorrelation functions $\mathcal{C}
(\phi^\circ)$ of the Power spectrum of the interfacial
deformations. The color density plot shows the evolution of the
autocorrelation function versus driving strength (left y-axes).
The quadratic order in the water sugar solution is weak and a
standard plot of the autocorrelation function at $a_0 = 25 m/s^2$
is shown as an example (black line, right y-axes). The lower fluid
layer is (a) the pure dyed water sugar solution and (b) with 110
ppm PEO. \label{fig5} }
\end{figure}

The melting processus can be characterized by the normalized
angular autocorrelation function $\mathcal{C} (\phi^\circ)$,
defined by van de Water and Binks \cite{binks97} such that square
pattern will lead to a pronounced peak at
$\mathcal{C}(\phi=90^\circ)$ (Fig. \ref{fig5}). For the polymer
solution a quadratic order is found up to driving strength $a_0
\approx 30 m/s^2$. For the water sugar solution the
$\mathcal{C}(\phi=90^\circ)$ peak is much weaker and disappears
for $a_0 > 20 m/s^2$.

\begin{figure}
\includegraphics[ width=1.1\linewidth]{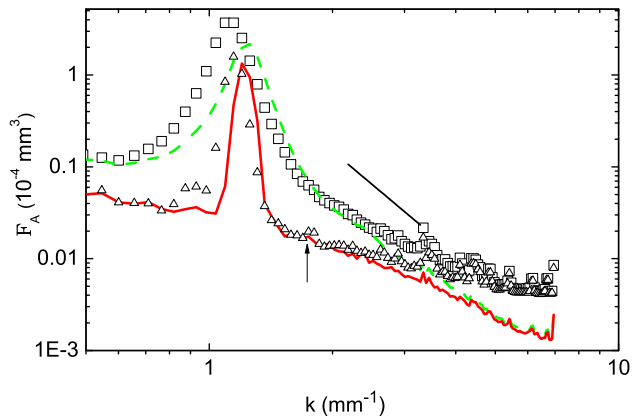}
\vspace{0.03cm} \caption{(Color online) The temporal and angular
averaged squared amplitude $\mathbf{F}_A$ of the Fourier spectrum
of the interfacial states versus the radial averaged wave number
$k$. Each curve is the average of 1000 images of the interfacial
wave state. The lower liquid is the water sugar solution (symbols)
or the 110 ppm PEO solution (lines). The driving amplitudes are
$a_0=31.7$ m/s$^2$ (triangles and red line) and $a_0=60$ m/s$^2$
(squares and green dashed line). The short black line indicates a
-15/4 slope. The arrow indicates the position of the first higher
order peak at $\sqrt{2} k_0$. \label{fig3}}
\end{figure}

Figure \ref{fig3} shows the temporal and angular averaged
amplitudes $\mathbf{F}_A$ of the Fourier spectrum. The prominent
peak at the basic wave number $k_0$ is clearly recognized. This is
the length scale at which energy is injected into the system.
However, already at intermediate driving strength $a_0=31.7$
m/s$^2$ the higher harmonic spatial Fourier peaks are very weak
and the spectrum almost consists of a continuum only. The theory
for weak turbulent capillary waves \cite{zakharov} predicts a wave
number density equivalent to
\begin{equation}
\mathbf{F}_A \sim k^{-15/4}.
\end{equation}
The exponent could be confirmed by experiments with a free surface
\cite{putterman} and our interfacial waves show the same
dependency at higher driving strengths, too (see straight line in
Fig. \ref{fig3}).

\begin{figure}
\includegraphics[ width=0.95\linewidth]{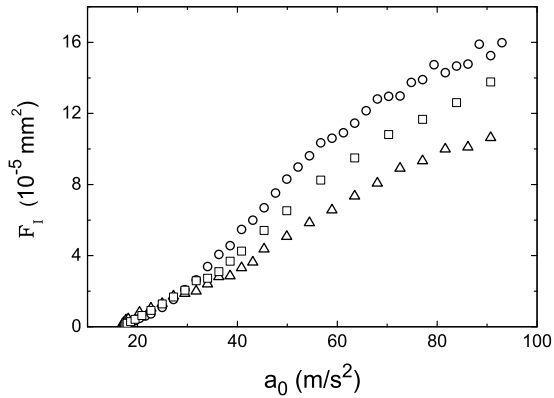}
\vspace{0.03cm} \caption{The integrated intensity $\mathbf{F}_I$
of the temporal and angular averaged spatial Fourier spectrum
(Fig. \ref{fig3}) of the interfacial states versus the driving
acceleration. The lower liquid is the water sugar solution
(circles), a 50 ppm PEO solution (squares) or the 110 ppm PEO
solution (triangles).\label{fig4} }
\end{figure}

We find a $~5\%$ difference in the basic wave vector $k_0$ for the
water and the polymer solution but for lower driving strengths the
main difference appears at large wave numbers. For the polymer
solutions the fluctuations in the small length scales - equivalent
to high temporal frequencies \cite{extended} - are almost one
order of magnitude lower than for the solution without polymers.
The integrated intensity $\mathbf{F}_I$ (see Fig. \ref{fig4}) is
dominated by the power of the basic wave vector and  is similar
for both systems at $a_0 \leq 40$ m/s$^2$ $\mathbf{F}_I$. This is
in contrast to earlier Faraday Experiments with more viscous
polymer solutions \cite{fauve}, where e.g. the critical
acceleration is significantly affected due to viscoelasticity.
Only for driving accelerations $a_0 \geq 40$ m/s$^2$ the
integrated intensities $\mathbf{F}_I$ of the solution with
polymers is lower than for the pure solution (see Fig. \ref{fig3}
and \ref{fig4}).

We think that the phenomena observed in our experiments can be
only interpreted if one takes into account the differences in
elongational and shear viscosity. A simple, Newtonian liquid is
defined by a constant viscosity $\eta$ that does not depend on the
applied shear rates or flow forms. Already in linear response the
viscosity of viscoelastic liquids becomes a complex quantity
$\eta^*(\omega)=\eta'(\omega)+i\eta''(\omega)$, with $\eta'$ the
dissipation and $\eta''$ the elasticity. It can be measured as
e.g. in an oscillatory shear experiment with small shear
amplitudes. In a similar experiment with constant shear rate
$\dot{\gamma}$ polymer solutions might show a shear thinning
behavior and $\eta(\dot{\gamma})$ depends on the shear rate. But
shear thinning is more pronounced for concentrated solutions. The
viscoelastic solutions used in the earlier Faraday Experiments
\cite{fauve} are characterized by a significant linear elastic
contribution $\eta''(\omega)$ and shear thinning. Their
interaction with the Faraday waves and the influence of the linear
stability could be well explained in terms of a linear response
theory.

The diluted low viscous polymer solution used in our experiments
does not show an elastic contribution $\eta'(\omega)$ that could
be measured with a high resolution rheometer (MARS, Thermo
Electron Germany, Karslruhe) and its shear viscosity measured at
different constant shear rates $1 < \dot{\gamma} < 100$ differs
only a few percent from the solvent viscosity. But diluted
solutions of flexible polymers are known to possess a pronounced
\textit{elongational viscosity} $\eta_e$ \cite{Amarouchene01}. Any
flow can be separated into a rotational and an elongational part.
This refers to the symmetric and the antisymmetric parts of the
velocity gradient tensor $\mathbf{L}$, with components
$L_{ij}=\nabla_j v_i$ ($v_i$ are the velocity components of the
flow). For Newtonian liquids $\eta_e$ is just given by a simple
geometrical factor, the Trouton ratio with $\eta_e = 3 \eta$. A
pure rotation has no effect on the polymers but an elongation
stretches them most efficiently at least when the Weisenberg
number Wi, the product of elongational rate $\dot{\epsilon}$ and
polymer relaxation time $\lambda$, exceeds a critical value $Wi=
\dot{\epsilon}\lambda > 1/2$ \cite{chu, relaxation}. In simple
shear flow the ratio between the symmetric and the antisymmetrical
parts of the velocity gradient tensor is one. The rotational part
leads to a tumbling movement of the polymers \cite{chu05} and the
stresses introduced by the elongational part are averaged over all
orientations. In pure elongational flow the polymer stress can
grow until the macroscopic elastic stress is orders of magnitude
larger than the viscous part. Consequently, the elongational
viscosities of our PEO solutions measured with a capillary break
up extensional rheometer (CABER, Thermo Electron Germany,
Karlsuhe) is in the order of several $Pas$, compared to $\eta_e =
3 \times 7.2 mPas$ for the solvent. Frequency dependent
measurements of the complex elongational viscosity $\eta_e^*$ do
not exist for our solutions, but for vanishing small amplitudes
near threshold it should not differ from the complex shear
viscosity $\eta^*$ in terms of a linear response theory.

We now can understand why the polymers do not affect the critical
accelerations: It is a particularity of surface (or interfacial)
waves that beside a small viscous sublayer near the interface the
flow profile can be described by a potential and the flow is
rotational free and purely elongational. The elastic part of the
linear complex viscosity $\eta''$ of our solutions is too small
and the dissipative part $\eta'$ does not differ from the solvent
viscosity. Only the contributions due to the pure elongational
character of the flow might affect the polymers, but this occurs
only for $Wi>1/2$.

To estimate when the flow is strong enough to affect the polymers
we look in a first approximation at only one spatial dimension.
The velocity potential for a given wave number $k$ has the form
\begin{equation}
\Phi = \omega /k A sin(kx)cosh(kz)cos(\omega t)
\end{equation}
with $A$ the deformation amplitude of the wave and $z$ the
direction of gravity. At the interface $z = 0$ the flow is
strongest and we can set $cosh(kz) = 1 $. It follows for the
amplitude $A_L$ of the velocity gradient tensor and thus the
maximum elongational rate that occurs at the position of maximum
peak elevation and at the interface $\dot{\epsilon}_{max} = A_L =
k \omega A$. Only at a driving strength of $a_0 = 40 m/s^2$ the
amplitude of the basic wave number is $A = 1 mm$ and the
elongational rate is $\dot{\epsilon} = 200$ 1/s and $Wi = 1.5$. At
this driving strength the amplitude of the basic mode is strong
enough to affect the polymers and we see differences in the
amplitude of $F_A$ at $k_0$, compared to the pure sugar water
solution. The integrated Intensity $F_I$ is dominated by the power
of the fundamental mode and the effect can bee seen best in Fig.
\ref{fig4}. However, already at low driving strength the wave
number continuum at higher wave numbers adds to the velocity
gradient tensor. The polymers get stretched by the high frequency
fluctuations and their amplitude is reduced (Fig. \ref{fig3}).
Apparently, these fluctuations are the reason for the disorder
transition.

A comparison with drag reduction in fully developed turbulence is
instructive. According to Landahl \cite{Landahl77} the occurence
of turbulent eddies in wall bounded systems like pipes is related
to bursting motions, the so-called streaks, regions of flow with a
high elongational component. The streaks occur near the wall and
interact subtle with the turbulent eddies in which the energy is
dissipated via a cascade to smaller lengths scales. The pronounced
elongational viscosity is supposed to inhibit the streaks,
yielding less eddies and less turbulent dissipation. Experimental
findings support this picture and even a direct correlation of
elongational viscosity and the ability of a polymer to perform to
turbulent drag reduction has been found\cite{Wagner02}.

In conclusion we have measured and described the effect of a small
amount of a flexible polymer on the transition from an ordered to
a weak turbulent state in parametrically driven interfacial waves.
Without polymers we observe a rapid transition from a square to a
disordered state with increasing driving strength. The addition of
the polymers suppresses the high frequency fluctuations and
stabilizes the ordered pattern. We find that the pronounced
elongational viscosity is responsible for this effect and we
suppose that further investigations on the effect of polymers on
weak turbulent flow might help us to understand some of the key
features of the scientific and technical important problem of
turbulent drag reduction, too.


\begin{references}
\bibitem{extended}
An extended presentation on the spatial and temporal spectra will
be published elsewhere.
\bibitem{relaxation}
The relaxation time $\lambda$ for the PEO in water was determined
by light scattering \cite{Amarouchene01} and assuming a linear
increase with the solvent viscosity we get $\lambda \approx 7$ ms.
Higher values have been reported too \cite{Wagner05}, but they
refer to large elongations.
\end{references}
\end{document}